\documentclass[aps,prl,reprint,groupedaddress]{revtex4-2}

\usepackage{aas_macros} 
\usepackage{graphicx}
\usepackage{amsmath}
\usepackage{xcolor}

\graphicspath{{./}{figs/}}

\begin{document}


\title{Physically Interpretable Machine Learning for nuclear masses}


\author{M.~R.~Mumpower}
\email[]{mumpower@lanl.gov}
\homepage[]{https://www.matthewmumpower.com}
\affiliation{Theoretical Division, Los Alamos National Laboratory, Los Alamos, NM, 87545, USA}

\author{T.~M.~Sprouse}
\affiliation{Theoretical Division, Los Alamos National Laboratory, Los Alamos, NM, 87545, USA}

\author{A.~E.~Lovell}
\affiliation{Theoretical Division, Los Alamos National Laboratory, Los Alamos, NM, 87545, USA}

\author{A.~T.~Mohan}
\affiliation{Computer, Computational and Statistical Sciences Division, Los Alamos National Laboratory, Los Alamos, NM, 87545, USA}


\date{\today}
\preprint{LA-UR-22-21855}

\begin{abstract}
We present a novel approach to modeling the ground state mass of atomic nuclei based directly on a probabilistic neural network constrained by relevant physics. 
Our Physically Interpretable Machine Learning (PIML) approach incorporates knowledge of physics by using a physically motivated feature space in addition to a soft physics constraint that is implemented as a penalty to the loss function. 
We train our PIML model on a random set of $\sim$20\% of the Atomic Mass Evaluation (AME) and predict the remaining $\sim$80\%. 
The success of our methodology is exhibited by the unprecedented $\sigma_{RMS}\sim186$ keV match to data for the training set and $\sigma_{RMS}\sim316$ keV for the entire AME with $Z \geq 20$. 
We show that our general methodology can be interpreted using feature importance. 
\end{abstract}


\maketitle

\textit{Introduction -} The minimal energy required to break up a nucleus into its constituent nucleons is one of the fundamental properties of an atomic nucleus. 
This quantity, which is equivalent to the mass, features prominently as an input for the theoretical prediction of a number of nuclear properties which are important for both scientific and technological applications \cite{Thoennessen+10, Talou2014, Kondev2021}. 
This effect is perhaps most apparent in the important role that masses play in predicting the reaction and decay properties of atomic nuclei \cite{Kawano2016, Randrup2021}. 
Masses also serve as critical inputs for the study of astrophysical phenomena, from influencing the composition of neutron star crusts to impacting heavy element synthesis and its potential observable consequences \cite{Steiner2012, Misch2013, Mumpower2016, Sprouse2020, Zhu2021}. 

The many-body Hamiltonian that describes atomic nuclei is exceedingly complex and remains unknown, therefore the lowest energy state cannot be calculated directly from first principles for heavy nuclei. 
This state of affairs has led to the development of many theoretical descriptions of atomic masses, including semi-classical approaches \cite{Moller2016}, microscopic approaches \cite{Goriely2009} and more recently, models enhanced by considering improvements to model discrepancies with data \cite{Utama2016, Neufcourt2019}. 
An inherent limitation in contemporary modeling is that the model itself remains fixed with optimization focused on parameters. 
This can be overcome with application of Machine Learning (ML) algorithms in which the model itself is optimized \cite{Wu2019}. 

In this letter we present a novel approach to modeling masses directly from a ML model constrained by physics. 
Our `Physically Interpretable Machine Learning' or PIML approach builds physically meaningful feature spaces and applies soft constraints to ensure relevant physics is being obeyed. 
It can be generalized to any problem in which physical constraints may need to be applied to a machine learned model. 
In a drastic improvement to our previous work \cite{Lovell2022}, we train a probabilistic network on a fraction of available data and predict the vast majority of masses for thousands of nuclear species measured to date. 
We achieve unprecedented model accuracy and retain predictive power when extrapolating. 
We show that our model can be interpreted using a standard measure for feature importance, and this interpretation fits within the context of the well-established picture of atomic nuclei. 

\textit{Methods - } We use a probabilistic ML technique, the Mixture Density Network \cite{Bishop1994}. 
Our Probabilistic network is built on the \texttt{PyTorch}~\cite{Paszke2019} framework and can be run on either CPU or GPU architectures. 
This type of modeling has been shown to be successful in describing nuclear properties while providing well-quantified uncertainties \cite{Lovell2020}. 

Lovell \textit{et al.}~\cite{Lovell2022} reported that a combination of macroscopic and microscopic features is suitable for describing masses across the chart of nuclides. 
Based on this previous analysis, we use eight features: the proton number ($Z$), the neutron number ($N$), the mass number ($A$), the odd-even nature of protons ($Z_\textrm{eo}$), the odd-even nature of neutrons ($N_\textrm{eo}$), the valence number of protons as measured from the nearest closed shell ($V_{p}$), the valence number of neutrons as measured from the nearest closed shell ($V_{n}$) and a measure of isospin asymmetry ($P_\textrm{asym}=\frac{N-Z}{A}$). 

The last five features inform the model on quantum mechanical effects.
Pairing effects manifest from the inclusion of the $Z_\textrm{eo}$ and $N_\textrm{eo}$ terms which are binary, taking the value of 0 or 1. 
Valence terms characterize the counting of particles (or holes) between major closed neutron and proton shells. 
As the valence number increases up to the mid-shell, more complex excitations, including collective modes may appear \cite{Casten1999}. 
The success of this picture can be related to nuclear promiscuity (a measure of the strength of proton-neutron interactions per valence nucleon) \cite{Casten1987}. 
The final feature informs the model about the Pauli exclusion principle. 

We take as our training set a random selection of the masses of 450 nuclei in the AME2016 \cite{Wang2017} with $Z \geq 20$. 
The same set of 450 nuclei is fixed throughout training and does not change. 
The match to this data is computed with a log loss function that we denote by $\mathcal{L}_{1}$ (see Ref.~\cite{Lovell2022}). 

In addition to the physics-based feature space, we seek to encode physical constraints into model training. 
For this work, we chose to enforce one such possibility, the Garvey-Kelson relations \cite{Garvey1969}. 
This well-known series of formulas involves a judicious choice of mass differences of neighboring nuclei that minimizes the interactions between nucleons to first order, resulting in particular linear combinations that strategically sum to zero. 
We implement this as a soft constraint, a second loss function, $\mathcal{L}_{2}$, in our training. 
This additional loss function is calculated over the entire AME and serves as a penalty for model solutions that do not obey well-established physical law. 
Note that because this is a soft constraint, training may ensue that temporarily increases $\mathcal{L}_{2}$ in pursuit of the global minimum. 
We revisit this important point shortly.

The model hyperparameters are as follows: the number of hidden layers is six, the number of hidden nodes is eight, the number of Gaussian ad-mixtures is one, the weight of the physics constraint is $\lambda_{\textrm{phys}}=1$, and we implement the Adam optimizer with initial learning rate 0.0002 \cite{Kingma2017}. 
To avoid overfitting, we implement regularization with a weight decay set to 0.01. 
These hyperparameters were determined from a select set of runs where the values were varied. 

The weight of the physics constraint, $\lambda_{\textrm{phys}}=1$ is especially noteworthy. 
We found that if the physics constraint was weighted too heavily (large values of $\lambda_{\textrm{phys}}$), training often failed as sharp cusps were encountered in the evolution of the total loss function which was prohibitive to optimization. 
As $\lambda_{\textrm{phys}}$ tends to zero, the physics constraint becomes less influential on training and we return to the previous results of Ref.~\cite{Lovell2022}. 

\begin{figure}
\includegraphics[width=\columnwidth]{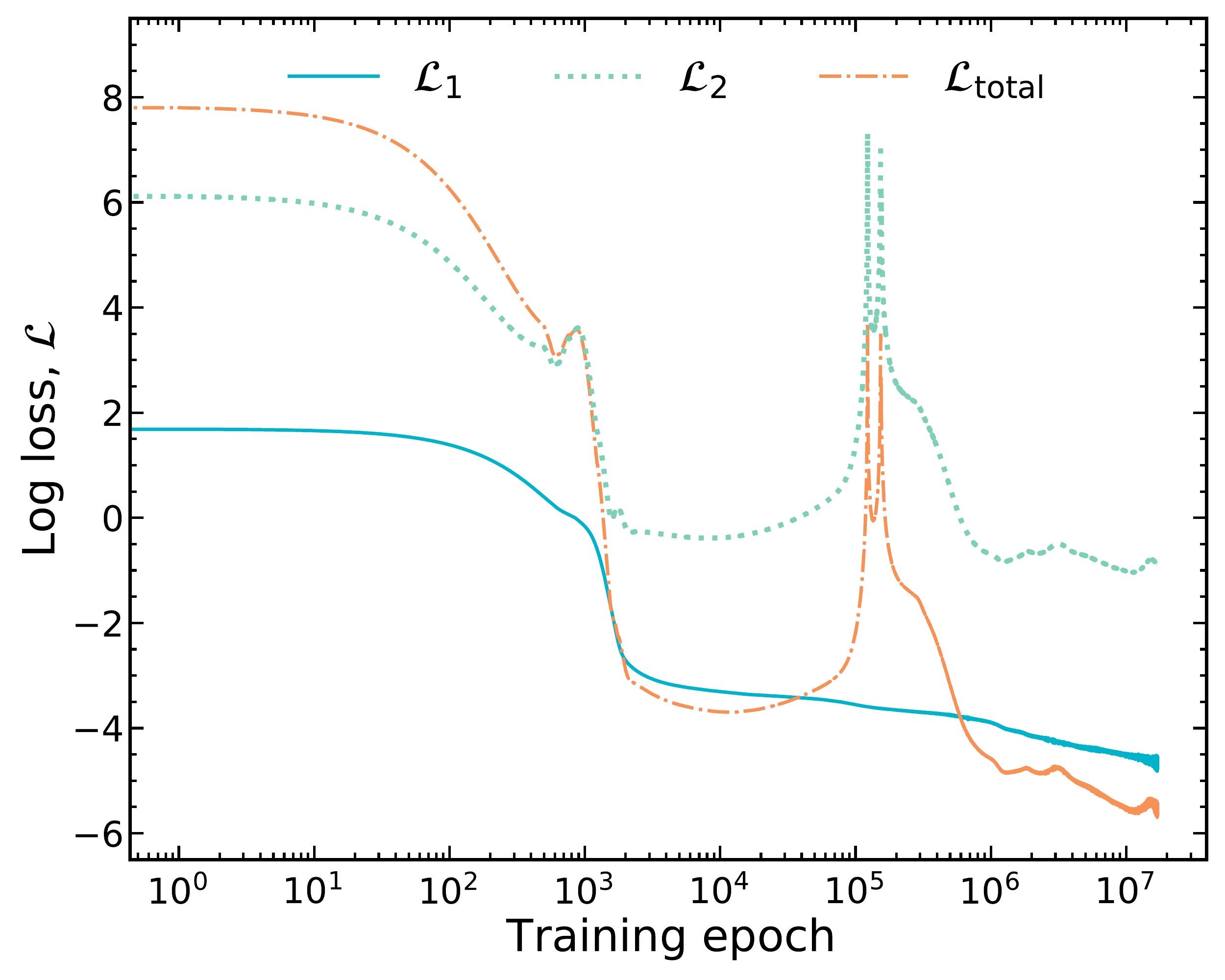}
\caption{The log loss as a function of training epoch for the match to data, $\mathcal{L}_{1}$, physics constraint, $\mathcal{L}_{2}$ and the sum of the two,  $\mathcal{L}_{\textrm{total}}=\mathcal{L}_{1}+\lambda_{\textrm{phys}}\mathcal{L}_{2}$. The learning rate is 0.0002. }
\label{fig:logloss}
\end{figure}

In training we seek to minimize the total log loss that consists of a sum of the loss for the match to our training set, as well as the physical constraint: $\mathcal{L}_{\textrm{total}}=\mathcal{L}_{1}+\lambda_{\textrm{phys}}\mathcal{L}_{2}$. 
Each training epoch attempts to improve the total log loss function with respect to previous solutions. 
We allow this process to continue for roughly $10^7$ epochs. 

The log losses for a sample training run used in this work is shown in Fig.~\ref{fig:logloss}. 
The loss with respect to data, $\mathcal{L}_{1}$, decreases monotonically by virtue of the minimization algorithm. 
The loss with respect to physics, $\mathcal{L}_{2}$, however, exhibits a complex and highly non-linear structure. 
This results in a total loss function that displays similarly complex behavior, e.g. briefly rising (ca. epoch no. $10^5$), before establishing a global minimum for the entire run. 

Figure~\ref{fig:logloss} demonstrates this very general but often overlooked attribute of optimization with multiple constraints \cite{Deb2005, Gunantara2018}. 
If we consider epoch numbers $\approx 1-8 \times 10^5$~, for example, we clearly see that while training based on fit to data \textit{alone} is generally quite good (and improving), the same fit's ability to satisfy basic physical requirements is quite poor by comparison. 
Global minima across all epochs, e.g.~in the case of a different training set, may not necessarily be the final epoch at which a stop condition is reached (although the two points do coincide by happenstance in Fig.~\ref{fig:logloss}). 
Consequently, it is important that past work that has applied ML in physics \textit{without implementing or considering physically motivated constraints} should be approached with caution. 

\begin{figure}
\includegraphics[width=\columnwidth]{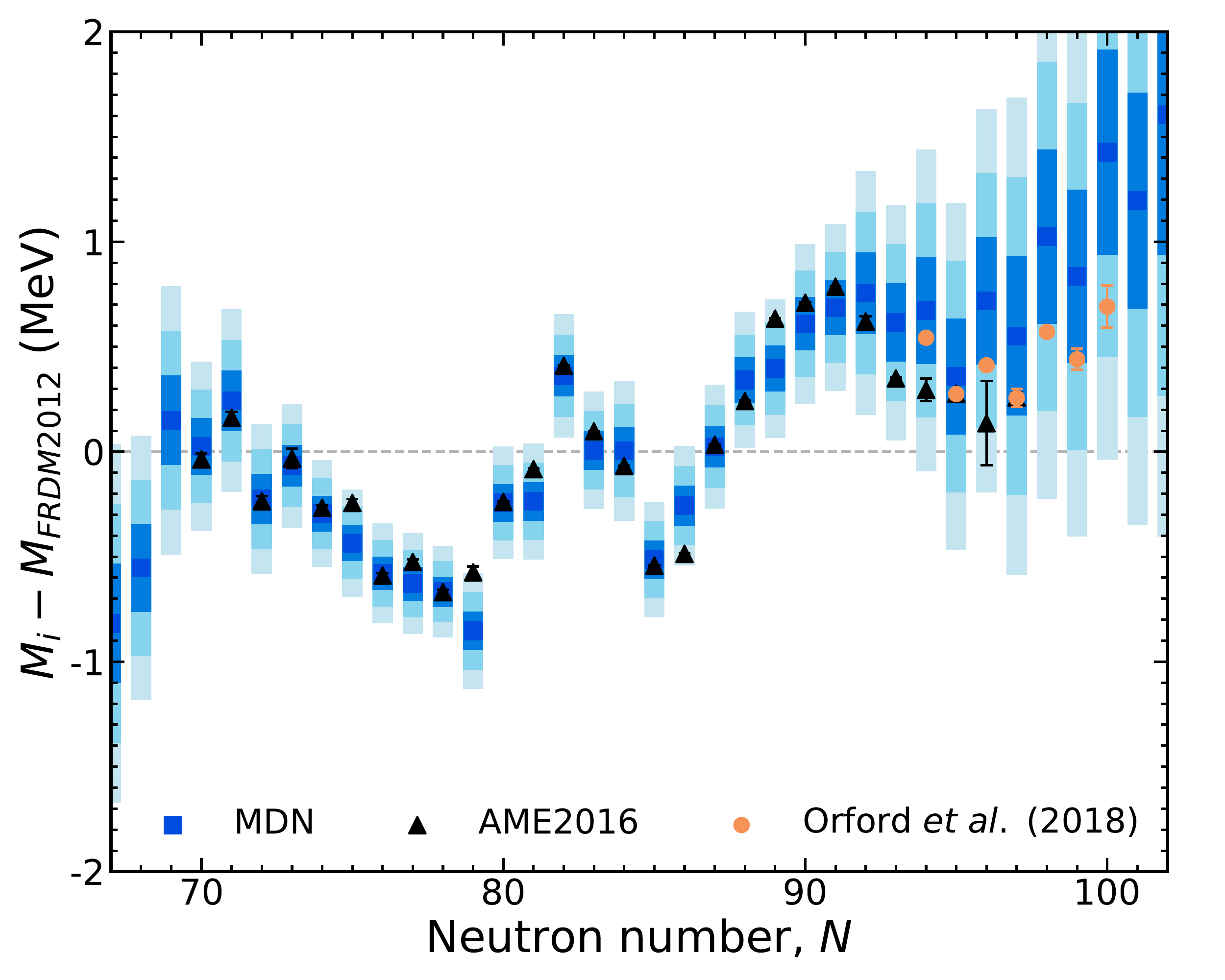}
\caption{Predictions of masses along the neodymium ($Z=60$) isotopic chain using our PIML model. The PIML average and 1-3$\sigma$ uncertainties are shown in progressively lighter shading along with AME2016 \cite{Wang2017} and recent data from Ref.~\cite{Orford2018}. Masses are plotted relative to FRDM2012. }
\label{fig:isochain_Z60}
\end{figure}

\textit{Results - } In Figure~\ref{fig:isochain_Z60}, we compare the PIML model predictions against our training data (AME2016 atomic masses) over a range of neodymium isotopes with existing measured data. 
The entirety of the AME dataset lies within the 3$\sigma$ uncertainty interval, with a majority (all but $\approx$ 3) data points lying within a 1$\sigma$ or 2$\sigma$ interval. 
The overall excellent agreement with AME data shown here is indicative of the results across the entire chart of nuclides. 
We report a root-mean-square error $\sigma_\textrm{RMS}$ for the PIML model to be ${\sigma_\textrm{RMS}\sim186}$~keV for the training set (20$\%$ coverage of AME2016 data) and ${\sigma_\textrm{RMS}\sim316}$~keV for the entirety of AME2016 with $Z \geq 20$. 
We retain our predictive capability (${\sigma_\textrm{RMS}\sim336}$~keV) when comparing to the latest 2020 release of the evaluation \cite{Wang2021}. 
These results are competitive with global mass models available today.  

Of primary interest to the nuclear physics community is not just the ability to model the properties of known nuclei (here, \textit{masses}), but also the ability to apply these same models to predict the properties which currently cannot be produced or otherwise studied in laboratory settings. 
This has, to date, been a particularly difficult situation for nuclear theory, because while, e.g., microscopic, macroscopic-microscopic, or phenomenological approaches to modeling atomic nuclei may be applied to nuclei well outside the range of those used for parameter calibrating, it has proven challenging to develop a robust picture of the overall uncertainty in the extrapolated predictions. 
Recent efforts have included the application of Gaussian processes and Bayesian methods to better quantify the uncertainties of model parameters, particularly those of microscopic nuclear models (see Ref.~\cite{Neufcourt2018, Melendez2019, Schunck2020} and references therein). 

Figure \ref{fig:isochain_Z60} gives some insight into how we may proceed with regards to understanding both the quality of PIML extrapolations beyond available data, as well as how these predictions can assist in our understanding of \textit{total} (statistical, systematic, and/or model) uncertainties of nuclear mass predictions. 
In particular we note that the results of our calculations were derived from both training and testing against AME2016 data. 
In the time that has since passed, there have been a number of experiments which have provided data that extends beyond the limits of this dataset, and in particular, the results of Ref.~\cite{Orford2018} extended the measurements of the neodymium isotopic chain shown in Fig.~\ref{fig:isochain_Z60} towards a series of more neutron-rich isotopes, up to atomic mass number $A=160$. 

On the question of extrapolation, we see that PIML predictions (which were constructed with no information concerning the Orford \textit{et al.} data) would appear to generally follow the trends in nuclear mass suggested by the additional measurements, and indeed lying well within the 2$\sigma$ uncertainty intervals. 
This suggests the reasonable expectation that the PIML model (or other models founded on similar principles) will reasonably extrapolate towards the more exotic, short-lived nuclei without any grotesque violation of basic nuclear physics principles, consistent with the design goals of our approach as laid out in the Methods section. 

Furthermore, the PIML approach naturally provides thoroughly robust estimates with respect to the uncertainties point-wise for each individual prediction, i.e., each individual mass prediction is associated with its own, uniquely inferred uncertainty. 
Indeed, the uncertainties shown in Fig.~\ref{fig:isochain_Z60} by the shaded bands clearly reflect their expected behavior, insofar as the error bands are narrowly focused about their mean (but not unreasonably so) where AME data exists, while as the predictions begin to drift past the limits of the training/testing dataset, the same uncertainty bands begin to diverge, up to about 3$\sigma \sim$1 MeV for the isotopes shown. 
This opens up the possibility for more thorough uncertainty quantification analyses based on this general approach, which we intend to explore in future works. 

\begin{figure}
\includegraphics[width=\columnwidth]{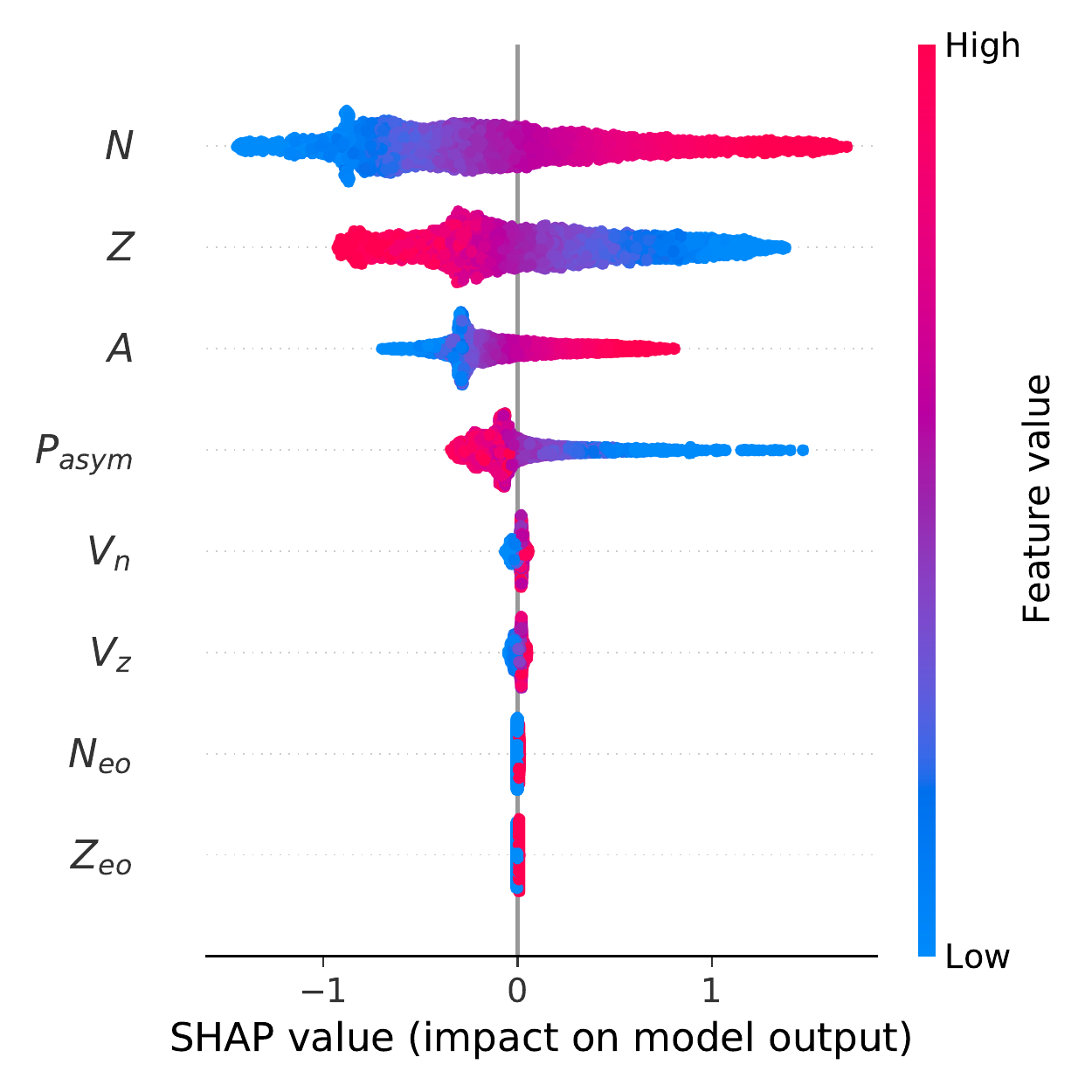}
\caption{The ranking of feature importance measured by SHAP value over the entire AME2020 \cite{Wang2021}. Macroscopic-like terms rank the highest followed by the quantum effects of Pauli exclusion, valence nucleons and pairing. }
\label{fig:shap_fi}
\end{figure}

An important consequence of PIML modeling is that the output of the network may be interpreted and understood, as with any useful theoretical model. 
To this end, we compute Shapley Additive Explanations (SHAP) values \cite{Lundberg+17} to measure feature importance. 
Figure \ref{fig:shap_fi} ranks the eight features when applied to the more recent AME2020 \cite{Wang2021}. 
We find that macroscopic quantities rank the highest in determining the masses, followed by the features which control the quantum effects. 
Figure \ref{fig:shap_fi} reflects the long held belief that, to first order, the atomic nucleus is well-described by bulk, macroscopic features, while microscopic features induce subtle, yet extremely important corrections \cite{Weizsacker1935, Negele1982}. 
This is evident when comparing a nucleus with its close neighbors, where the macroscopic features may be very nearly equal, but the quantum effects in these nuclei can lead to dramatic differences in nearly every nuclear observable. 

\textit{Conclusions -} We present a feed-forward neural network with the capacity to directly predict the nuclear binding of atomic nuclei. 
We achieve an unprecedented match to training data with a root-mean-square error of 186 keV ($\sim$ 20\% of the AME) while utilizing a single physical constraint and only eight parameters defining the feature space. 
Our model is capable of predicting all of the AME with $Z \geq 20$ at ${\sigma_\textrm{RMS}\sim316}$ keV. 

In contrast to previous Machine Learning methods, our Physically Interpretable Machine Learning (PIML) approach affords the ability to analyze manifestations of physical phenomena. 
We demonstrate this by establishing relative importance of macroscopic and microscopic input features which is consistent with the traditional understanding of the atomic nuclear binding energy viz. the semi-empirical mass formula, as well as with macroscopic-microscopic and fully microscopic methods. 

The technology developed here is general, and may in principle be applied to the study of any physical observable, or combination thereof. 
A concerted effort in this approach opens new possibilities to capture complex physics that is not currently viable via other contemporary approaches to computational physics. 
This may prove especially valuable in advancing the study of a wide range of many-body problems that permeate physics. 

\bibliography{refs.bib}


\section{Acknowledgements}\label{sec:ack}
We thank P.~Talou for his helpful comments and review of manuscript. 
M.R.M., T.M.S., A.E.L., A.T.M. were supported by the US Department of Energy through the Los Alamos National Laboratory (LANL). LANL is operated by Triad National Security, LLC, for the National Nuclear Security Administration of U.S.\ Department of Energy (Contract No.\ 89233218CNA000001). 

\end{document}